\documentclass[article,superscriptaddress,showpacs,floatfix]{revtex4}
\usepackage{amssymb,amsmath,epsfig,lscape,color,eucal,graphicx,bm,epsfig,multirow}

\usepackage[ruled,vlined]{algorithm2e} 
\usepackage{amstext,amssymb,amsfonts}
\usepackage{graphicx}
\usepackage{color}
\usepackage{epstopdf}
\usepackage{subfigure}
\usepackage[section]{placeins}
\usepackage{float}

\begin{document}

\title{Method for estimating critical exponents in percolation processes with low sampling}

\author{N.Bastas}
\affiliation{Department of Physics, University of Thessaloniki,54124 Thessaloniki, Greece}
\author{K.Kosmidis}
\affiliation{Department of Physics, University of Thessaloniki,54124 Thessaloniki, Greece}
\affiliation {School of Engineering and Science,Jacobs University Bremen, Campus Ring 1, 28759 Bremen, Germany.}
\author{P. Giazitzidis}
\affiliation{Department of Physics, University of Thessaloniki,54124 Thessaloniki, Greece}
\author{M.Maragakis}
\affiliation{Department of Physics, University of Thessaloniki,54124 Thessaloniki, Greece}
\affiliation{Department of Economics, University of Macedonia, Thessaloniki, Greece}

\date{\today}

\begin{abstract}

In phase transition phenomena, the estimation of the critical point is crucial for the calculation of the various critical exponents and the determination of the universality class they belong to. However, this is not an easy task, since a huge amount of realizations is needed to eliminate the noise in the data. In this paper, we introduce a novel method for the simultaneous estimation of the critical point $p_c$ and the critical exponent $\beta/\nu$, applied for the case of ``explosive'' bond percolation on $2D$ square lattices and ER networks. The results show that with only a few hundred of realizations, it is possible to acquire accurate values for these quantities. Guidelines are given at the end for the applicability of the method to other cases as well.  

\end{abstract}


\maketitle

\section{Introduction}
\label{Intro}
Percolation is a well-known geometrical phase transition model, initially developed to describe the flow in porous media \cite{Hammersley}. Since then, the core idea has been applied to numerous other problems \cite{Stauffer_Book}, and has been extended by several other variants (continuum percolation, site-bond percolation, bootstrap percolation, invasion percolation, etc. \cite{Bunde_book}). 

Among them, a recent one that raised a debate and attracted considerable interest, is one which was termed ``explosive'' percolation \cite{Achlioptas2009a}. According to this model, we randomly pick $2$ candidate edges that do not yet exist, and keep the one which leads to the smallest cluster product, discarding the second. In this way, there is a delay in the formation of the largest component, accompanied with an abrupt transition. The abruptness of the transition had initially been explained as a discontinuous transition \cite{Friedman2009,Cho2009,Ziff2010a,Radicchi2010a}. However, subsequent studies and rigorous results have established guidlines for ``achlioptaslike'' processes to be considered either continuous or discontinuous \cite{daCosta2010,Riordan2011a,Cellai2011,Nagler2012,Riordan2012,LiuYY2012,Cho2013,Schroder2013,Oliveira2014}. Moreover, various models have been proposed which exhibit a truly discontinuous transition \cite{Rozenfeld2010,Schrenk2011a,Chen2011a,Schrenk2012a,Boettcher2012,Chen2013b,Sigh2014,Chae2014,Buldyrev2010,LiWei2012,Gao2013}. For more information on the work related to ``explosive percolation'', see \cite{Bastas_review}.

An important thing in phase transition phenomena is to identify the nature of the change. The universality class of a specific transition is one of the most important indicators on this issue. By determining the values of the critical exponents at the transition point, one is able to define the universality class the system under study belongs to. Such quantities are computationally demanding, and thus difficult to calculate. In order to maintain good statistical ensembles and calculate accurately the critical point and the exponents associated to certain quantities, a large amount of individual realizations is usually performed. As an example, in \cite{Ziff2010a}, at least $150000$ simulations of a system of size $2048\times2048$, and up to $1000000$ for smaller system sizes were performed.

This paper, presents a method which calculates the critical exponents of a percolation process with equal, or even better accuracy, but with lower statistical sampling. More specifically, we develop a simple tweak of a previously presented method found in \cite{bastas2011explosive}. By doing a slight modification in it, we are able to determine simultaneously $p_c$ and the critical exponent $\beta/\nu$, with significant accuracy, from data averaged over a very small amount of realizations (of the order of $10^2-10^3$). Using these values we obtain also the $1/\nu$ exponent and thus define the universality class of the system. We apply this method to a hybrid percolation model which is essentially the same to that described in \cite{Liu2012,Fan2012}. This specific model exhibits an interesting behavior, useful to display the power of the suggested method, as it interpolates between the Achioptas process and random percolation.

The rest of the paper is organized as follows: in section \ref{method} we illustrate the method. In section \ref{ResANDdis}, we apply this method to a hybrid bond percolation model in both lattices and networks and extract the critical threshold and critical exponents. Finally, in section \ref{conc} we summarize our results. The algorithm of the method is given in the Appendix.

\section{Method}
\label{method}

In order to accurately calculate the critical exponents one must first estimate with high precision the percolation threshold $p_c$ of the system. There are many ways to find $p_c$, e.g. wrapping-around probabilities, ``reduced'' second moment or susceptibility peak scaling \cite{Ziff2010a}, second largest cluster peak scaling \cite{Yi2013} or even power-law fitting of the cluster size distribution versus system size, $L$, for different $p$ \cite{Radicchi2010a}. As soon as the critical point is correctly estimated, one  calculates the quantity of interest at $p_c$ and performs power-law fit on a logarithmic scale. The slope found corresponds to the relevant critical exponent (e.g. the exponent $\beta/\nu$ is derived from $P_{max}$ versus $L$).

An inadequate number of realizations can lead to a problem in defining the critical point and subsequently the exponents. In such cases with high noise, a scientist must decide whether the crossing point is clear enough or not, or whether a curve is a straight line or not. By doing a large number individual runs, one minimizes (but never ultimately removes) the possibility to have significant errors with respect to the number of runs. 

In \cite{bastas2011explosive}, a new method was proposed, based on the scaling relation of the form:

\begin{equation}
\label{eq1}
X = L^{-x} F[(p-p_c)L^{1/\nu}]
\end{equation}
near $p_c$ [$X$ is an arbitrary quantity of interest (e.g. size of the largest cluster $S_{max}$, susceptibility $\chi$), $x$ is a critical exponent characterizing this quantity and $1/\nu$ is the correlation length exponent]. Let us define $Y=XL^x$. When $p = p_c$, eq. (\ref{eq1}) suggests that all systems (independent of the system size $L$) have a crossing point, since $X=F(0)L^{-x}$ and thus $Y=F(0)$. The pairwise differences, $\Lambda(p,x)$, of $Y$ for different systems, $i$ with size $L_i$ and $j$ with size $L_j$, are specified as:

\begin{equation}
\label{eq2}
\Lambda(p;x) = \sum_{i\neq j}{(Y_{L_i}-Y_{L_j})^2}
\end{equation}
This means that in theory $\Lambda=0$ at $(p_c;\beta/\nu$). In practice, due to the inherent noise of simulations, one should expect that $\Lambda(p,x)$ acquires its minimum for the previous set of values. 

In a similar manner, if we take $1/Y$ and insert this in eq. (\ref{eq2}), it is expected that $\Lambda$ acquires its minimum for the same set of parameters. The combination of the two approaches leads to the equation: $H_{L_i}=Y_{L_i} + 1/Y_{L_i}$. This equation presents again a crossing point at $p = p_c$. In this case, the minimization function acquires the form:
 
\begin{equation}
\label{eq3}
\Lambda(p;x) = \sum_{i \neq j}{(H_{L_i}-H_{L_j})^2}
\end{equation}
It must be stressed that the method requires the simultaneous minimization of the square differences with respect to both the parameters $p$ and $\beta/\nu$. In order to derive the minimum value of the function of eq. (\ref{eq3}), the use of all available systems sizes ($L$ for lattices or $N$ for networks) is mandatory.

In the next section, we provide evidence for the validity of our method in determining the percolation threshold, $p_c$, and a critical exponent, $\beta/\nu$, estimating $1/\nu$, and comparing our results to those estimated by previous works. For a description of the procedure, which can be automated, refer to the Appendix.

\section{Results and Discussion}
\label{ResANDdis}

\begin{figure}[h!]
\centering 
\subfigure[]{
\includegraphics[width=5.5cm] {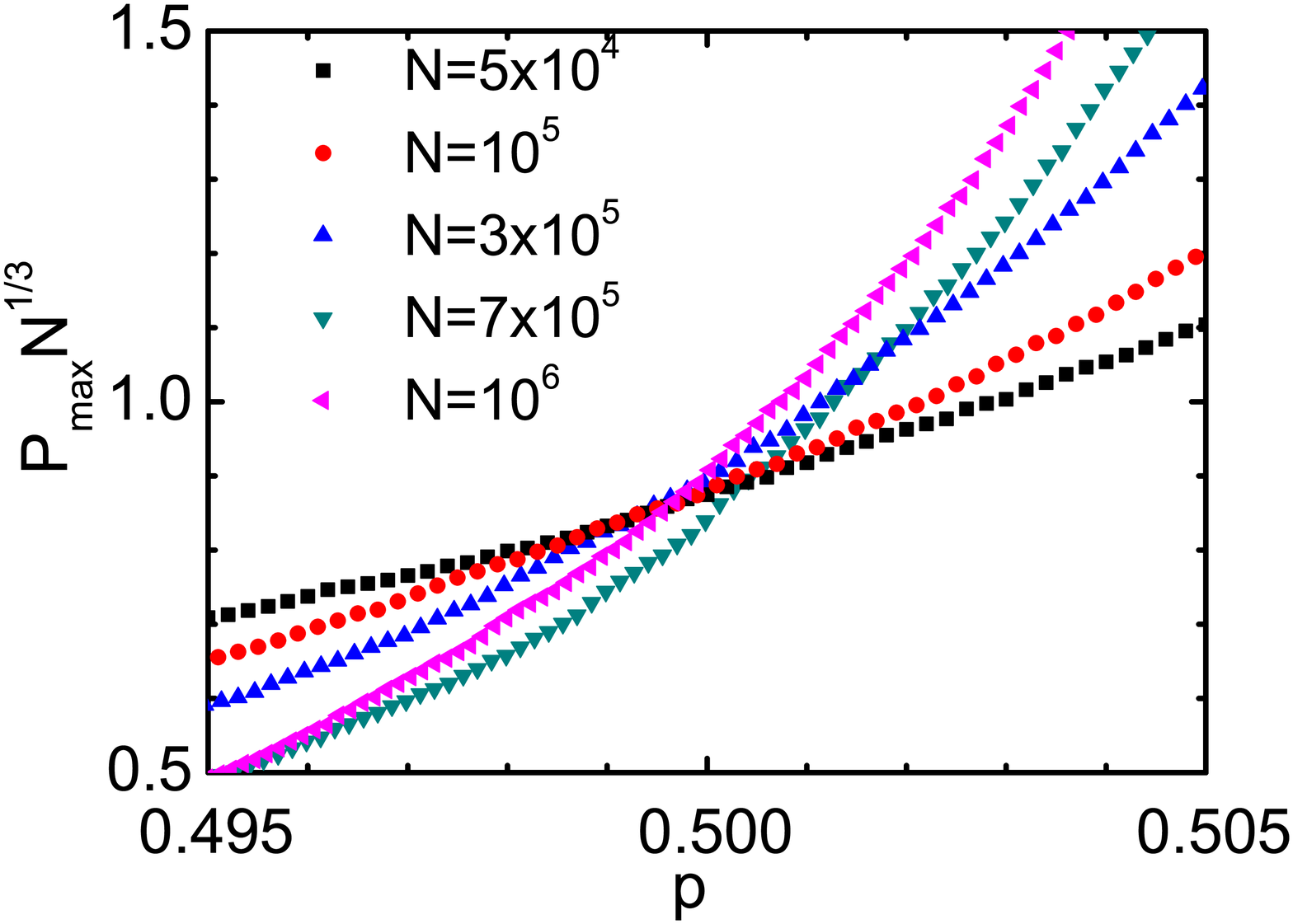} \label{Figure1a} }
\subfigure[]{
\includegraphics[width=5cm] {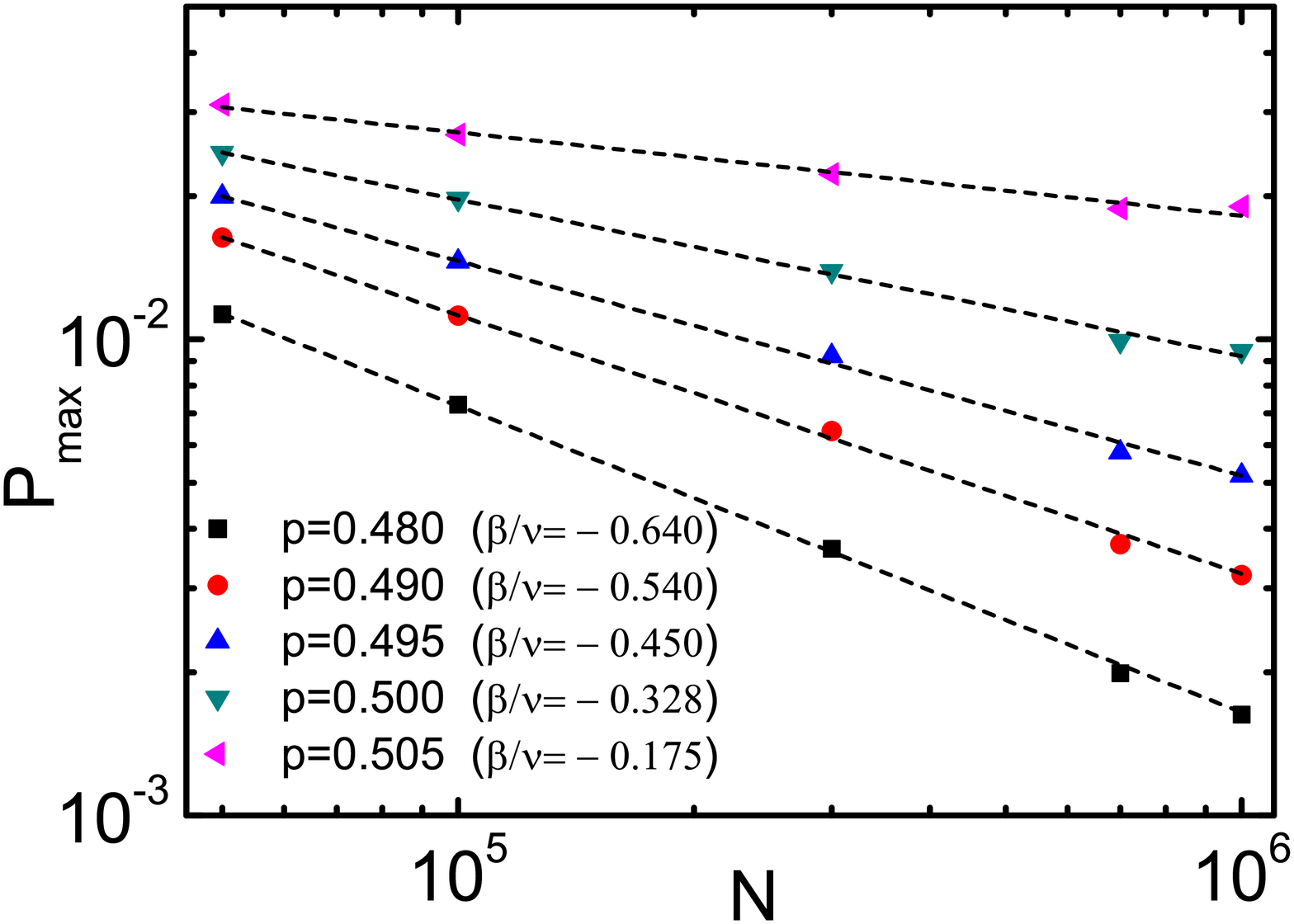} \label{Figure1b} }
\subfigure[]{
\includegraphics[width=5.5cm] {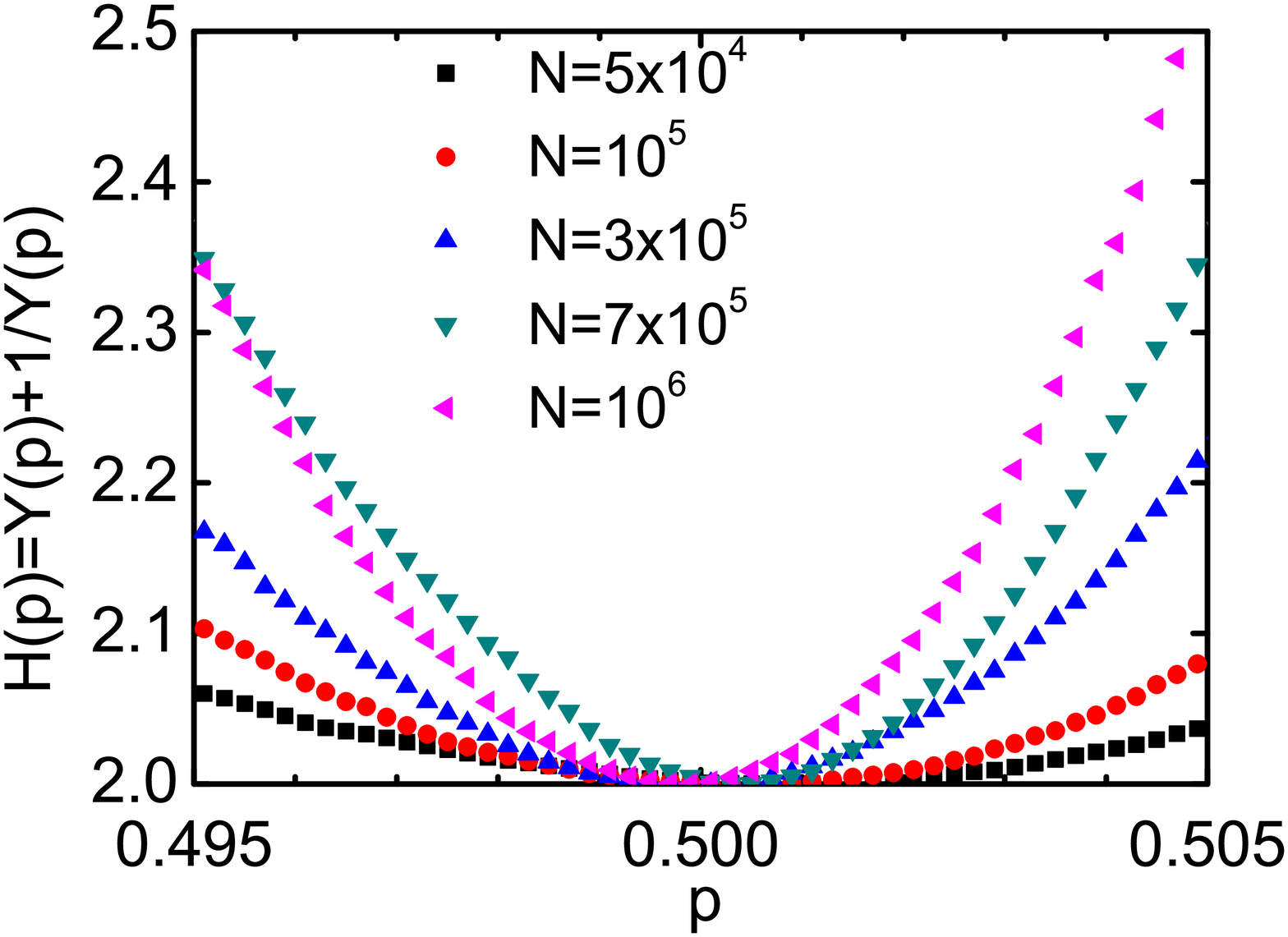} \label{Figure1c} }
\caption{(Color Online). Demonstration of the common methods used to estimate $p_c$ and $\beta/\nu$ for random percolation process on ER networks : (a) plot of $P_{max}N^{\beta/\nu}$ as a function of $p$ for the case of the theoretically expected value $\beta/\nu=1/3$. Different data plots do not cross at the same $p$ value as expected from scaling theory. (b) Plot of $P_{max}$ vs $N$ on log-log scale for $5$ values of $p$: $0.480$ (black squares), $0.490$ (red circles), $0.495$ (upward blue triangles), $0.500$ (downward dark cyan triangles) and $0.505$ (leftward magenta triangles). Dashed lines are power-law fittings on data. It is evident that the best choice is for $p=0.495$, which yelds $\beta/\nu =-0.45$. Both values are far from the theoretically expected. (c) Demonstration of the current method. By applying eq. (\ref{eq3}) we derive a percolation threshold value $p_c=0.5001$, and $\beta/\nu=0.338$. Symbols for (a) and (c) are: $N=50000$ (black squares),$N=100000$ (red circles), $N=300000$ (upward blue triangles), $N=700000$ (downward dark cyan triangles) and $N=1000000$ (leftward magenta triangles). $H$ is described in the text.} 
\label{Figure1}
\end{figure}

In order to demonstrate the validity and the advantages of the proposed method, we apply it in comparison with the most commonly used methods, in the case of ER networks for classic (ordinary) percolation. According to theory, for this system the percolation threshold is expected at $p_c=0.5$, and $\beta/\nu=1/3$. We will use a small amount of realizations ($500$). In Fig. \ref{Figure1a}, we plot $P_{max}N^{\beta/\nu}$ as a function of $p$, for the theoretically expected value of $\beta/\nu=1/3$. According to finite-size scaling arguments, one should expect that all the data plots cross at the same value which is $p_c=0.5$ (see eq.(\ref{eq1})). It is evident that this is not the case, as there is no such single point. In Fig.\ref{Figure1b}, we plot on a log-log scale $P_{max}$ as a function of the system size $N$ for $5$ different values of $p$: $0.480$, $0.490$ , $0.495$, $0.500$ and $0.505$. We perform the standard procedure of power-law fitting on the data points. The best fit line, according to theory \cite{Stauffer_Book}, is received when we are at $p_c$. In Fig. \ref{Figure1b}, this value is $p=0.495$, for which $\beta/\nu =-0.45$. Both values are far from the theoretically expected ones. Thus, for the amount of realizations used, we cannot estimate $p_c$ with good agreement. On the other hand, with the same set of data, the method proposed (see Fig.\ref{Figure1c}) can accurately estimate both $p_c$ and $\beta/\nu$. In this case, we get $p_c=0.5001$ and $\beta/\nu=0.338$, very close to the theoretically expected values. Similar results are obtained for the Achlioptas (explosive percolation) process. Thus, our method seems valid for use with low amount of data.

In \cite{Liu2012,Fan2012}, the authors propose a way to produce a hybrid model. At each time step, a link is placed between two nodes (or sites) either by the Achlioptas (product rule) process (with probability $1-q$) or in random (with probability $q$). Thus, for $q=0$, we recover the Achlioptas process \cite{Achlioptas2009a} and for $q=1$ the random percolation process \cite{Stauffer_Book}. The use of  this model enables us to demonstrate the applicability and strength of our method to predict percolation thresholds and exponents of unknown theoretical values. The simulations were performed on $2D$ square lattices for sizes $L=1000,1250,1500,1750,2000$ and for ER networks with $N=50000,100000,300000,700000$ and $1000000$ nodes. The averages were taken over only $1000$ realizations for the former and for $500$ realizations for the latter. We note here that despite the fact that it is common practice in the literature, our results were not convoluted, either at the individual run level or in the averaging level. This means that noise is present in our results, but as proven below, it does not significantly affect them.
 
\begin{figure}[h!]
\centering 
\subfigure[]{
\includegraphics[width=6cm] {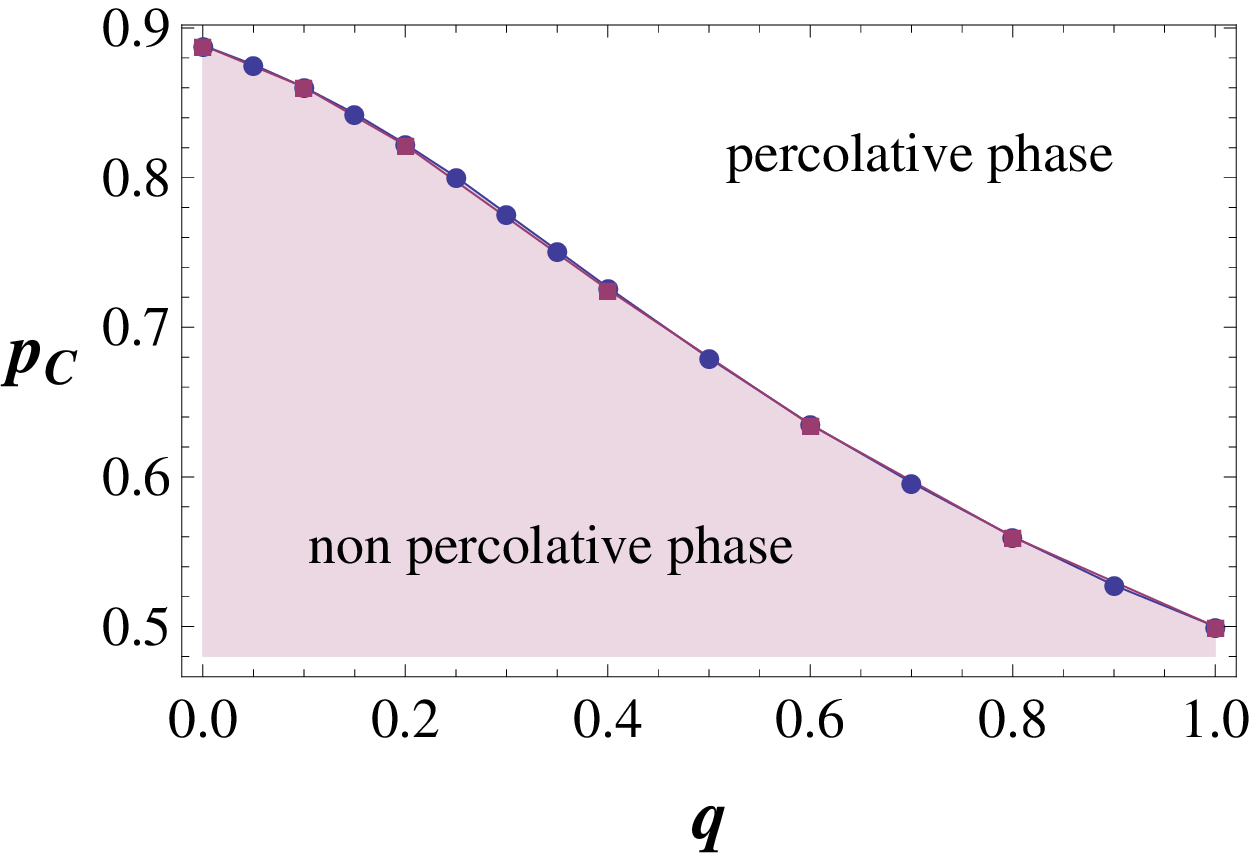} \label{Figure2a} }
\subfigure[]{
\includegraphics[width=6cm] {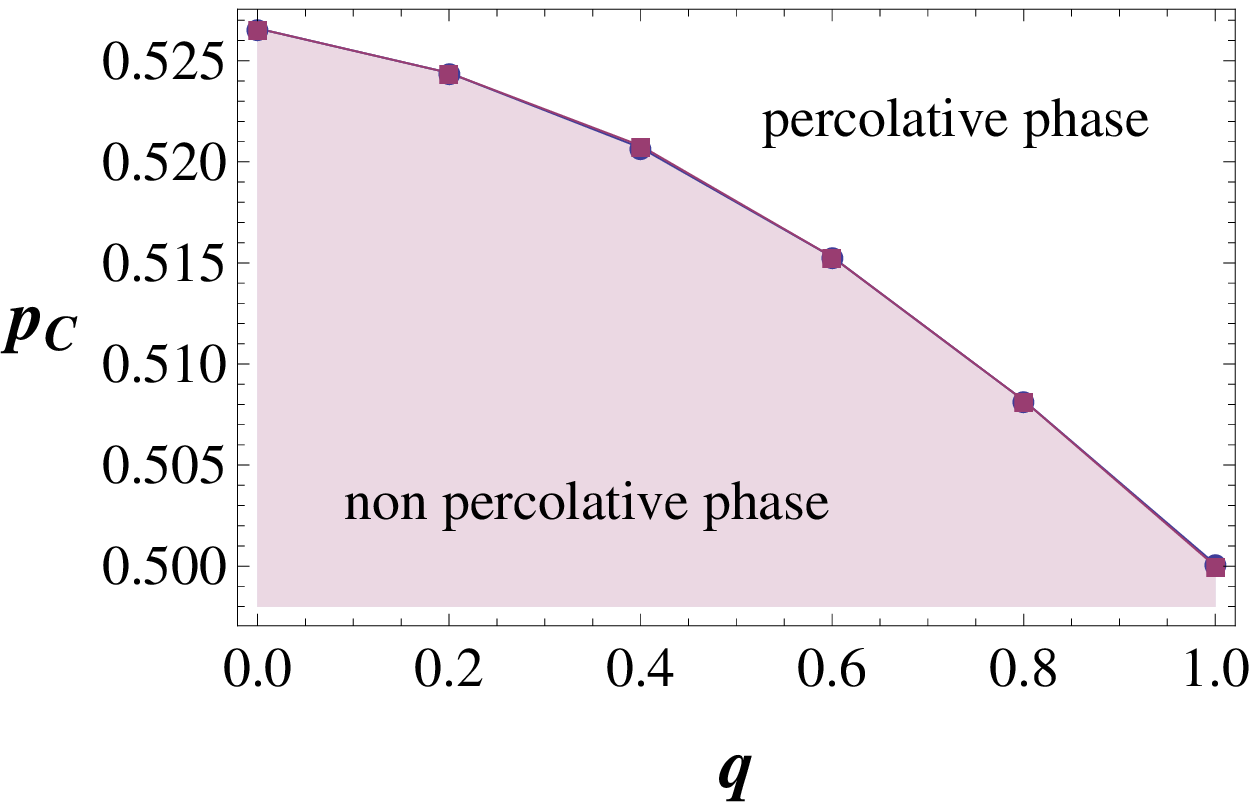} \label{Figure2b} }
\caption{(Color Online). Plot of $p_c$ as a function of $q$ for (a) ER networks and (b) $2D$ square lattices (circular symbols). For comparison, data from \cite{Liu2012,Fan2012} are also plotted (square symbols).} 
\label{Figure2}
\end{figure}

In Fig.\ref{Figure2}, we present the evolution of $p_c$ as a function of $q$, for $2D$ square lattices and ER networks. For comparison, we also plot the values estimated in \cite{Liu2012,Fan2012}. Even though we use data averaged over realizations $2-4$ orders of magnitude less than in \cite{Liu2012,Fan2012}, the values are almost identical to those predicted there (values of $q$ correspond to values of $p$ as it is defined in \cite{Liu2012,Fan2012}).

\begin{figure}[!]
\centering 
\subfigure[]{
\includegraphics[width=6cm] {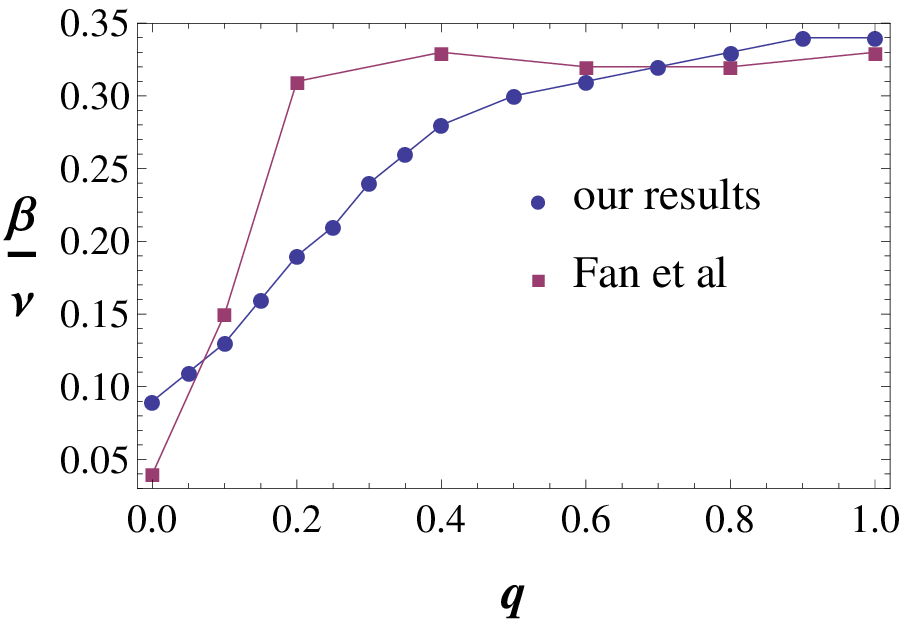} \label{Figure3a} }
\subfigure[]{
\includegraphics[width=6cm] {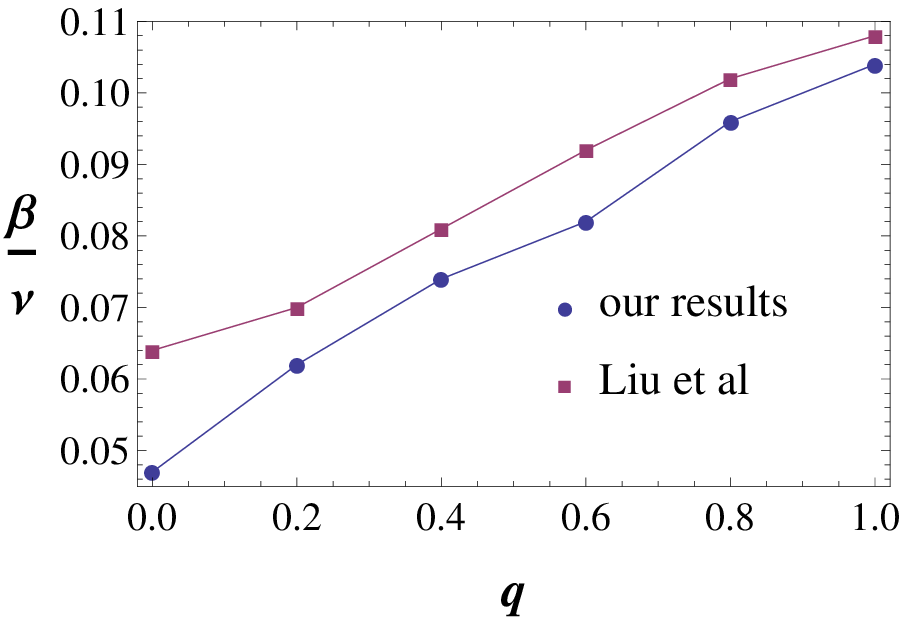} \label{Figure3b} }
\caption{(Color Online). Plot of $\beta/\nu$ as a function of $q$ for (a) ER networks and (b) $2D$ square lattices (circular symbols). For comparison, data from \cite{Liu2012,Fan2012} are also plotted (square symbols).} 
\label{Figure3}
\end{figure}

However, the case is different for the values of $\beta/\nu$. As it is shown in Fig. \ref{Figure3a}, our results indicate a gradual change for different values of $q$, while an abrupt transition is implied from the values in \cite{Fan2012}. In $2D$ square lattices (Fig. \ref{Figure3b}), both set of results are in good agreement. 

\begin{figure}[!]
\centering 
\subfigure[]{
\includegraphics[width=6cm] {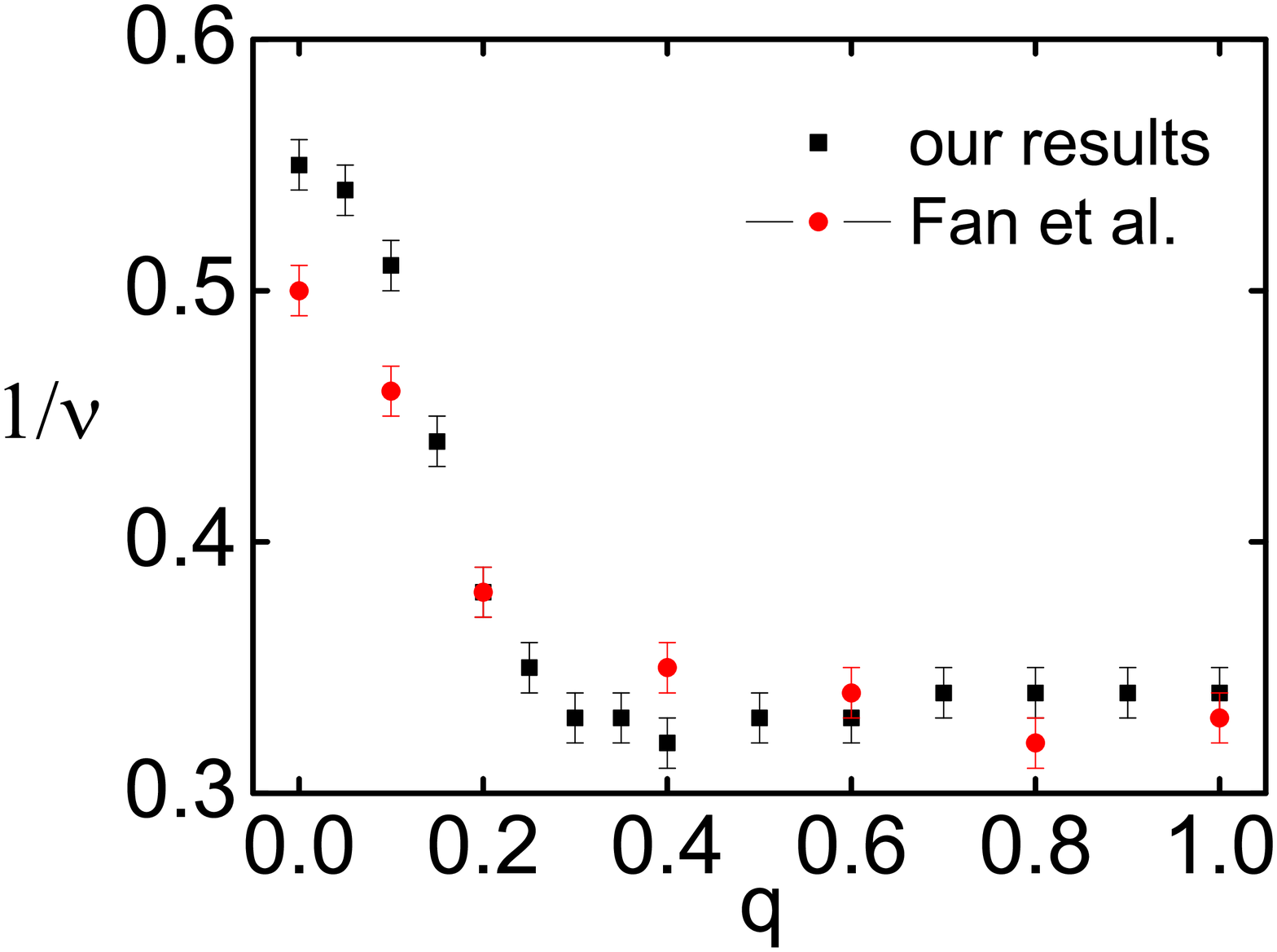} \label{Figure4a} }
\subfigure[]{
\includegraphics[width=6cm] {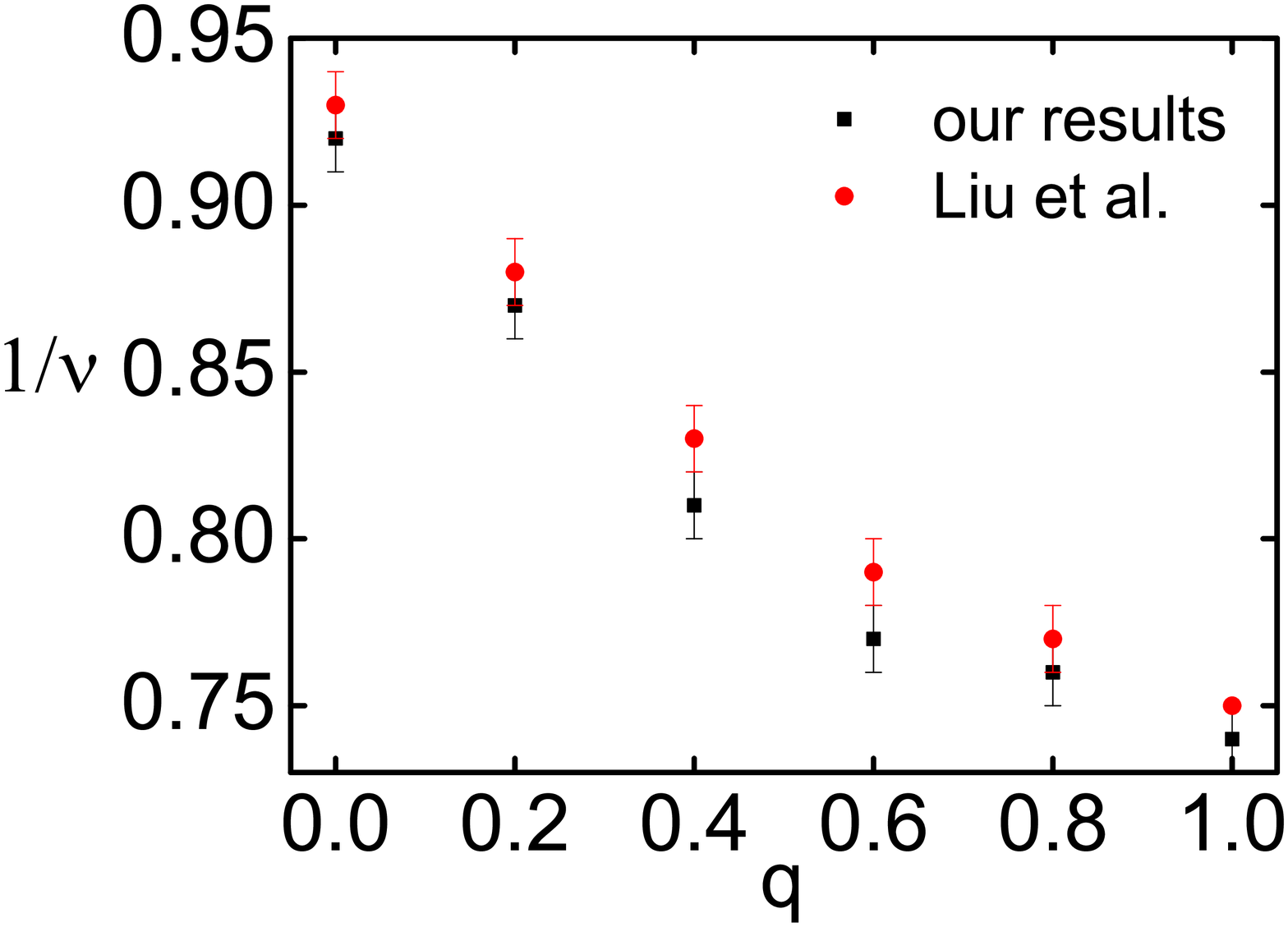} \label{Figure4b} }
\caption{(Color Online).Plot of $1/\nu$ as a function of $q$ for (a) ER networks and (b) $2D$ square lattices (square symbols). For comparison, data from references \cite{Liu2012,Fan2012} are also plotted with circular symbols.}  
\label{Figure4}
\end{figure}

The same behavior is also observed in Fig. \ref{Figure4}. Again, there are discrepancies between our results and the ones of reference \cite{Fan2012} for the case of ER networks, but $1/\nu$ behaves in the same manner as in reference \cite{Liu2012}.

It should be noted here that our method implies that there is a gradual change in the universality class, since both $\beta/\nu$ and $1/\nu$ gradually change from $0$ to $1$. In contrast, results from \cite{Fan2012} point to a steady value for both exponents from $q\approx0.4$ to $q=1$, which can translate to a single universality class for this entire region of $q$ values. This requires the oddity of a transition between a region of many continuously changing universality classes (from $q=0$ to $q\approx0.4$), and a second region (from $q\approx0.4$ to $q=1$) with only one universality class. We see no reason for such a transition. Instead, we believe that the results we get from our method with much less runs provide the more reasonable explanation of a smooth transition in the universality classes.

\begin{figure}[!]
\centering 
\includegraphics[width=6cm] {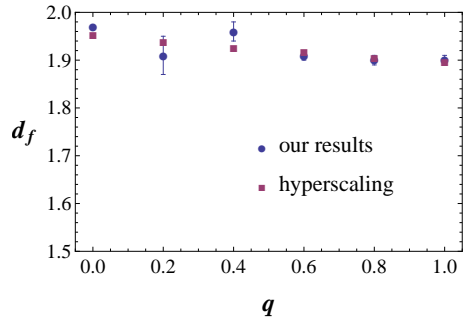}
\caption{(Color Online). Plot of $d_f$ as a function of $q$ for $2D$ square lattices (circular symbols). For comparison, we also plot the values calculated using the hyperscaling relation $d_f=d-\beta/\nu$ (square symbols). } 
\label{Figure5}
\end{figure}

Finally, for the case of $2D$ square lattices, we estimate the fractal dimension $d_f$ from the slope of the power-law fitting on the values of $S_{max}$ at $p_c$ as a function of $L$. In order to verify whether our results are valid, we extract the values of $d_f$, as calculated by the hyperscaling relation $d_f=d-\beta/\nu$. In Fig.\ref{Figure5}, it is evident that the hyperscaling relation is valid for each value of $q$ with satisfying accuracy. A similar hyperscaling relation does not hold for ER networks and thus our results there cannot be checked. Yet, our finding of the gradual change in the universality classes seems to validate our results.

\section{Conclusions}
\label{conc}

In this paper, we have developed a novel technique to significantly improve the reliability of the percolation threshold, $p_c$, and critical exponent, $\beta/\nu$, calculated values. The method requires a relatively simple reprocessing of the results extracted by any model. It can be automated (see Appendix) and thus it does not require any human interference. Moreover, it does not require a huge amount of realizations (commonly of the order of $10^6$ which are usually convoluted) but gives accurate results for realizations of the order of $10^3$ (which do not even have to be convoluted). It also allows for the determination of very small exponent values, such as the ones found in explosive percolation, with low computational cost. Finally, the method proposed in \cite{bastas2011explosive}, was used in \cite{Chen2013b} to define the critical point $p_c$ for the second giant component in the BFW model. The method presented in this paper is an extension of the previous method. Thus, it is probable that this method is suitable for other models as well.

\section{Acknowledgments}
This work used the European Grid Infrastructure (EGI) through the National Grid Infrastructures NGI\_GRNET, HellasGRID as part of the SEE Virtual Organization.
N. B. acknowledges financial support from Public Benefit Foundation Alexander S. Onassis. 

\section{Appendix}

Here, we present the automated method used to define the percolation threshold $p_c$ and the critical exponent $\beta/\nu$ in a pseudocode form.
\\

\begin{algorithm}[h!]
		\SetKwInOut{Input}{input}\SetKwInOut{Output}{output}
		\SetInd{1em}{0em}
		\BlankLine
		\Input{Datasets in the form ($links$,$S_{max}$)}
		\Output{$p_c$ and $\beta/\nu$ }
		\BlankLine
		\Begin{
			\Indp read datasets\;
			\BlankLine
			translate datasets in the form ($p$,$P_{max}N^{a}$)\;
			\BlankLine
			Define interpolation function $Y$ on the translated datasets \;
			\BlankLine
			Define function $H=Y+1/Y$ \;
			\BlankLine
			Form the minimization function $\Lambda(p;x)$ \;
			\BlankLine
			\SetInd{1.5em}{1em}
			\For{$p = p_{lower}$ \KwTo $p_{upper}$}{
					\BlankLine
					Set $\Lambda(p;x)$\;
					\BlankLine
					minimize $\Lambda(p;x)$ with respect to $x$\;
					\BlankLine
					\SetInd{1.5em}{1em}
					\If{$\Lambda$ is global minimum} {store $(p,x)$}
					
				}
				\BlankLine
				$p_c = p$\;
				$\beta/\nu = x$\;
		}
		\BlankLine
		Output $(p_c,\beta/\nu)$ \;
		\BlankLine
		Perform Collapse of $H$ vs $L$ $\rightarrow$ estimate $1/\nu$\;
		\BlankLine
		
		\caption{Defining $p_c$ and $\beta/\nu$}
\end{algorithm}

\bibliographystyle{unsrt}
\bibliography{explosivePercolation}

\end{document}